\documentclass[aps,pre,superscriptaddress,amssymb,twocolumn,showpacs]{revtex4}
\usepackage{graphicx}
\usepackage{epsfig}
\usepackage{graphics}
   \usepackage{setspace}

\begin{document}

\title{Photonic gaps  in cholesteric elastomers under deformation}

\author{P.~Cicuta}
\affiliation{Cavendish Laboratory, University of Cambridge,
Madingley Road, Cambridge, CB3 0HE, U.K. }
\affiliation{Nanoscience Center, University of Cambridge,
J.J.Thomson Avenue, Cambridge, CB3 0FF, U.K. }
\author{A.R.~Tajbakhsh}
\affiliation{Cavendish Laboratory, University of Cambridge,
Madingley Road, Cambridge, CB3 0HE, U.K. }
\author{E.M.~Terentjev}
\affiliation{Cavendish Laboratory, University of Cambridge,
Madingley Road, Cambridge, CB3 0HE, U.K. }


\begin{abstract}
Cholesteric liquid crystal elastomers have interesting and
potentially very useful photonic properties. In an ideal
monodomain configuration of these materials, one finds a
Bragg-reflection of light in a narrow wavelength range and a
particular circular polarization. This is due to the periodic
structure of the material along one dimension. In many practical
cases, the cholesteric rubber possesses a sufficient degree of
quenched disorder, which makes the selective reflection broadband.
We investigate experimentally the problem of how the transmittance
of light is affected by mechanical deformation of the elastomer,
and the relation to changes in liquid crystalline structure. We
explore a series of samples which have been synthesized with
photonic stop-gaps across the visible range. This allows us to
compare results with detailed theoretical predictions regarding
the   evolution of stop-gaps  in cholesteric elastomers.
\end{abstract}


\maketitle

\section{Introduction}
In liquid crystal elastomers, the rubbery polymer network is made
of chains that include  rod-like mesogenic units as sections of
the main chain or as side chains. The physical properties and the
response of these materials to external fields are extremely rich
and  lead to a variety of applications. Experiments and
theoretical studies in this field have been carried out by a
number of groups for some time and this topic is broadly reviewed
in the recent monograph \cite{WT03}.

\begin{figure}
\resizebox{0.4\textwidth}{!}{\includegraphics{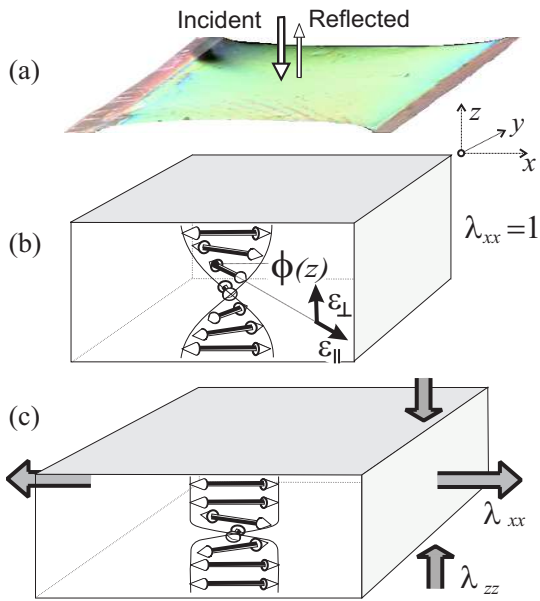}}
\caption{(color online)(a)~Photograph of a CLCE film, of size
approximately 1$\times$1cm$^2$, under small extension. The green
color is given by selective reflection caused by the periodic
structure. (b)~Diagram sketching the CLCE geometry of a sample at
rest, $\lambda_{xx}=1$. At $\lambda_{xx}=1$ the cholesteric
helical texture is characterized by a uniform director rotation
about the pitch axis $z$. (c)~Under uniaxial strain
$\lambda=\lambda_{xx}$ imposed along $x$ the helical structure
initially coarsens, leading to regions of faster and slower
director rotation. The associated contraction of sample thickness
is $\lambda_{zz}$. In all the experiments  incident light is
parallel to the pitch axis.} \label{diagram}
\end{figure}

A cholesteric (chiral nematic) phase can be induced in a nematic
liquid crystal by adding of small quantity of chiral molecules,
called the chiral dopant in this context. In an ideal cholesteric
elastomer (CLCE) the liquid crystal structure is locally nematic,
and a director can be identified in each $x$-$y$ plane forming an
angle $\phi$ to the $x$~axis  \cite{deGennes93}. The director
rotates continuously as a function of~$z$, see Fig.~\ref{diagram},
forming a periodic helical structure characterized by a pitch $p$
(or equivalently by the wave~number $q=2\pi/p$). Because of the
local quadrupolar symmetry of the nematic order, the periodicity
interval along $z$ is only $p/2$. Cholesteric liquid crystals have
complex optical properties due to their high local birefringence
and the modulation on the length scale $p$, often comparable with
the wavelength of light \cite{Belyakov79}. For example, circularly
polarized light with  wavelength $\Lambda_{0}$ incident on the
sample will not be transmitted if its handedness is the same as
the cholesteric helix and its wavelength is $\Lambda_{0}= p \,
\bar{m}$, where $\bar{m}=\frac{1}{2}(m_{\rm o} + m_{\rm e})$ and
$m_{\rm o}$ and $m_{\rm e}$ are the indexes of refraction of the
liquid crystal. This determines a photonic stop-band, that is a
1D-bandgap, for a narrow range of wavelengths and one circular
polarization of light. Furthermore, linearly polarized light on
either side of the bandgap is very strongly rotated.

The study of advanced optical properties of CLCE has only become
possible recently, in spite of the fact that the materials were
first synthesized a long time ago \cite{Finkelmann:81}. The
breakthrough has been achieved by Kim and Finkelmann
\cite{Kim:01}, who developed a technique of forming monodomain
cholesteric structures crosslinked into the rubbery network; in
most other circumstances a network ends up polydomain, with strong
quenched disorder -- it scatters light and is not useful for any
optical purpose. Advantages of having a well-ordered material with
a photonic stop-gap that can be altered by mechanical deformation
of the rubber have been demonstrated by, e.g. the use of CLCE as
cavities in tunable lasers \cite{Palffy-Muhoray:01}. Further
applications will emerge as the surprising properties of these
materials become better known.

\begin{figure}
\resizebox{0.4\textwidth}{!}{\includegraphics{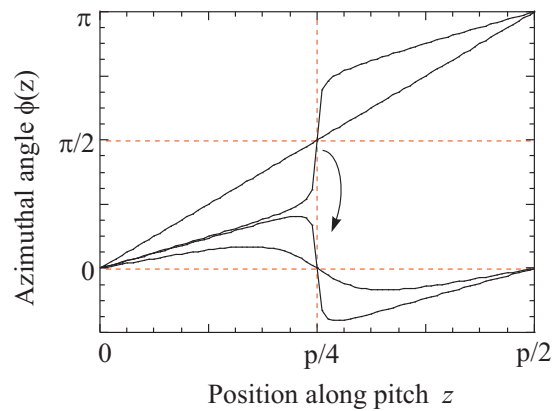}}
\caption{The director distribution $\phi$ inside the elastomer,
calculated  from Eq.~(\ref{phi}). The initially uniform helix
coarsens as the sample is strained, as indicated by the arrow. At
a critical strain $\lambda_{c}$ the texture becomes non-helical.
See Ref.~\cite{Warner:00} for details.} \label{theocoarse}
\end{figure}

In recent work \cite{Cicuta:02} we have studied the elastic
behavior of a monodomain CLCE and its relation to the underlying
structure, through a combination of structural (X-ray) and
mechanical probes. Optical investigation was also employed to
provide precise measurements of the sample deformation.  In brief,
the behavior under uniaxial strain is complex, because the
director tends to align along the stress axis but this is resisted
by the anchoring of the local director to the initial crosslinked
helical texture. Theoretical analysis \cite{Warner:00} predicts in
detail how the ideal cholesteric texture is modified, as soon as
an external strain $\lambda=\lambda_{xx}$ is applied. The director
angle $\phi$ is given by  the formula:
\begin{eqnarray}
&&\tan \, 2\phi = \label{phi} \\
&&=\frac{2\lambda^{1/4} (r-1) \sin \, 2 \tilde{q}z}
{(r-1)(\lambda^2+\lambda^{-3/2})\cos \,2 \tilde{q}z
+(r+1)(\lambda^2- \lambda^{-3/2}) },\nonumber
\end{eqnarray}
where $\lambda$ stands for the imposed strain $\lambda_{xx}$ and
the parameter $r$ is the intrinsic measure of local polymer chain
anisotropy induced by the local uniaxial anisotropy of the
mesophase (see \cite{WT03} for detail). To simplify the results of
\cite{Warner:00}, in the Eq.~(\ref{phi}) we have approximated the
transverse contraction $\lambda_{zz}$ (due to the rubber
incompressibility) by $\lambda^{-1/4}$, and the resulting affine
contraction of the cholesteric pitch is taken into account by
defining a reduced wave~number $\tilde{q}=\lambda^{1/4}q$.
Eq.~(\ref{phi}) expresses the strain-induced bias of director
orientations along the $x$-direction. The initially uniform helix
coarsens continuously until a critical strain $\lambda_{c}$ is
reached, corresponding to the point of zero denominator in the
right hand side of Eq.~\ref{phi}. At this point, the coarsened
helical director texture discontinuously jumps into a non-helical
state, modulated along $z$-axis but with the azimuthal angle
$\phi(z)$ not acquiring any phase, instead undulating back and
forth about the $x$-direction. This behavior is shown in
Figure~\ref{theocoarse}.
The critical strain can be easily estimated as $\lambda_{c} =
r^{2/7}$, from  Eq.~(\ref{phi}). The parameter of backbone chain
anisotropy, $r$, is the single parameter of the  theory of ideal
nematic (and cholesteric) elastomers; $r=\ell_\|/\ell_\bot$ is
interpreted as the average ratio of backbone chain step lengths
along and perpendicular to the local director, or
$r=R^2_\|/R^2_\bot$ is the squared ratio of principal radii of
polymer chain gyration. In nematic elastomers, $r$ can be directly
measured by examining the uniaxial expansion on cooling the
elastomer, thus increasing the magnitude of the nematic order
parameter. In a monodomain cholesteric rubber one can see a
similar thermal phenomenon, represented as a uniform biaxial
extension in the $x$-$y$~plane on cooling the CLCE, cf.
Fig.~\ref{diagram}. From such a measurement reported in
\cite{Kim:01} we can estimate that for this class of materials $r
\approx 1.16$, leading to $\lambda_{c} \simeq 1.04$ (that is, a
critical strain of 4\%). Recently much attention has been paid to
evaluating and interpreting this parameter of anisotropic (locally
uniaxial) networks \cite{GreveFW,Clarke01}.

Eq.~(\ref{phi}) was tested in \cite{Cicuta:02} and its key
predictions have been verified. In particular, by measuring the
non-trivial  contraction in the $z$-direction it was  seen that
the deformation in the $x$-direction is, in fact, semi-soft. It
was also confirmed that the average of director orientation under
strain corresponds to the distribution expected from
Eq.~(\ref{phi}). The coarsening of the helix has been predicted to
produce complex non-linear corrections to the optical response of
the material, which are very different from the effect of an
external field on a conventional cholesteric liquid phase
\cite{Bermel:02a,Bermel:02}.

In this paper we study the rich optical properties of monodomain
cholesteric elastomers. A range of samples is investigated that
have been synthesized with stop-gaps across the visible range. We
show how transmission and polarization of light can be controlled
and manipulated through mechanical deformation of rubber films, by
exploiting  the coupling of the locally birefringent liquid
crystalline texture to the underlying polymer network. We compare
our results with the predictions of~\cite{Bermel:02a}.

\section{Methods}
\subsection{Synthesis of materials}

Free standing strips of single-crystal and well aligned
cholesteric rubber have been prepared following the general
principle suggested by Kim and Finkelmann \cite{Kim:01}. A
uniaxial deswelling is induced in a weakly crosslinked and swollen
cholesteric gel, laterally constrained in a centrifuged film. The
reduction of thickness while constraining the lateral dimensions
is equivalent to a symmetric biaxial extension in the plane of the
film. This  leads to the macroscopic orientation of the director
in the plane. Therefore, the  helical axis is  aligned
perpendicular to the plane of the polymer film. This orientation
is then locked by completing a second-stage cross-linking of the
rubbery network. We describe the materials in a reference system
with the $z$~axis parallel to the helical axis, as shown on  the
diagram in Fig.~\ref{diagram}.

\begin{figure} 
\resizebox{0.45\textwidth}{!}{\includegraphics{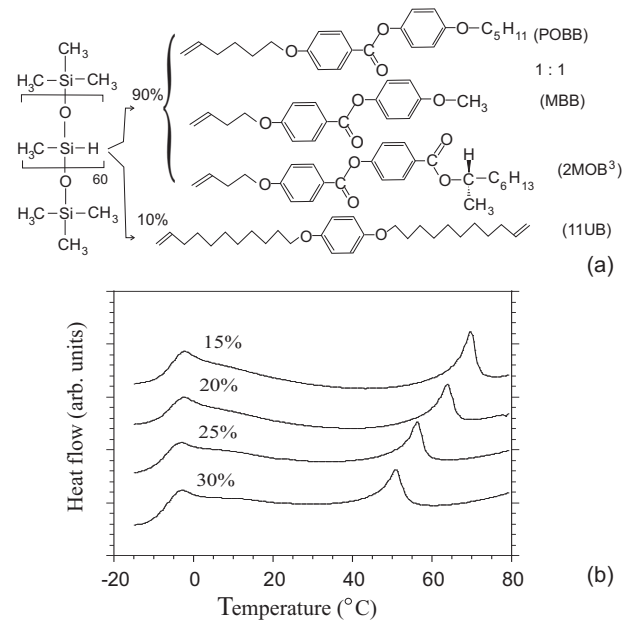}}
 \caption{(a)~Chemical structures of the compounds forming the
cholesteric elastomers under investigation. A siloxane backbone
chain is reacted with 90\,mol\% mesogenic side groups and
10\,mol\% of the flexible di-functional crosslinking groups
(11UB). The rod-like mesogenic (nematic) moieties are divided in
the proportion 1:1 between POBB and MBB; the chiral $R(-)$
rod-like moiety 2MOB${}^3$ has been taken in the proportion of 15,
20, 25 and 30\,mol\% with respect to the nematic groups.
(b)~Differential scanning calorimetry results for the four
materials studied in this work, indicating the nematic
(cholesteric) clearing point and the glass transition variation
with network composition. \label{chem}}
\end{figure}

New side-chain cholesteric liquid crystal elastomers are
synthesized in-house following the general method summarized
above. Polysiloxane backbone chains are reacted with 90\,mol\%
mesogenic side groups and 10\,mol\% of the two-functional
crosslinker 1,4\,di(11-undeceneoxy)benzene, labelled as 11UB in
Fig.~\ref{chem}(a). The choice and the composition of mesogenic
groups, see Fig.~\ref{chem}(a), is dictated by a number of
considerations. First of all, we mixed two different nematogenic
groups, 4-pentyloxyphenyl-4'-(4''-buteneoxy) benzoate (POBB) and
4-methoxyphenyl-4'-(4''-buteneoxy) benzoate (MBB), in equal
proportion. This gives a strong nematic phase with no
crystallization tendency and a low glass transition. The chirality
is brought to the system by including the $R(-)$ 4-[(2
methyloctyl) benzoate]-4'-(4''-buteneoxy) benzoate (2MOB${}^3$) at
different concentrations.

A word on terminology is due here. There are several ways of
describing the chiral substances, developed in chemistry. The old
Rosanoff (1906) notation distinguishes between D[+] (for {\it
dextra}) and L[--] (for {\it laevo}) on the basis of relative
arrangement four different bonds of the chiral carbon. The
Cahn-Ingold-Prelog (1956) notation is also based on ranking of
bonds according to specific sequence rules, so that it can be used
for more complex molecules; it specifies R[+] (for {\it rectus},
clockwise rotation) and S[--] (for {\it sinister}, anticlockwise).
It is only natural, that different sources of organic chemistry
data have mixed notations, e.g. the Aldrich catalogue quotes
chiral molecules as R-(--) and S-(+). This simply reflects the
fact that different chirality indices (scalar and tensorial) must
be introduced to describe different physical responses, while the
proper notation is not yet developed in spite of many recent
advances \cite{Osipov95,Harris99}. For instance, the sense of
steric chirality (asymmetry in the geometric shape of the object)
is not necessarily the same as that of the third-order dielectric
polarizability $\beta_{ijk}$ (determined by electronic structure)
and that, in turn, may be different at different frequencies, for
instance for the rotation of light polarization. The macroscopic
``phase chirality'' of cholesteric materials is a result of
cooperative action of all such effects and is simply distinguished
 by the circular dichroism, so that a right-handed (clockwise,
R*) material reflects right-hand polarized
light~\cite{deGennes93}.

The technique of preparing monodomain cholesteric rubber requires
that the crosslinking is carried out during slow de-swelling in
the cholesteric phase and so it was essential that our mesogenic
groups form the phase on their own -- before polymerization. The
mixture of the chosen monomers is a room-temperature cholesteric,
achieving this purpose.

\begin{figure*} 
\resizebox{0.43\textwidth}{!}{\includegraphics{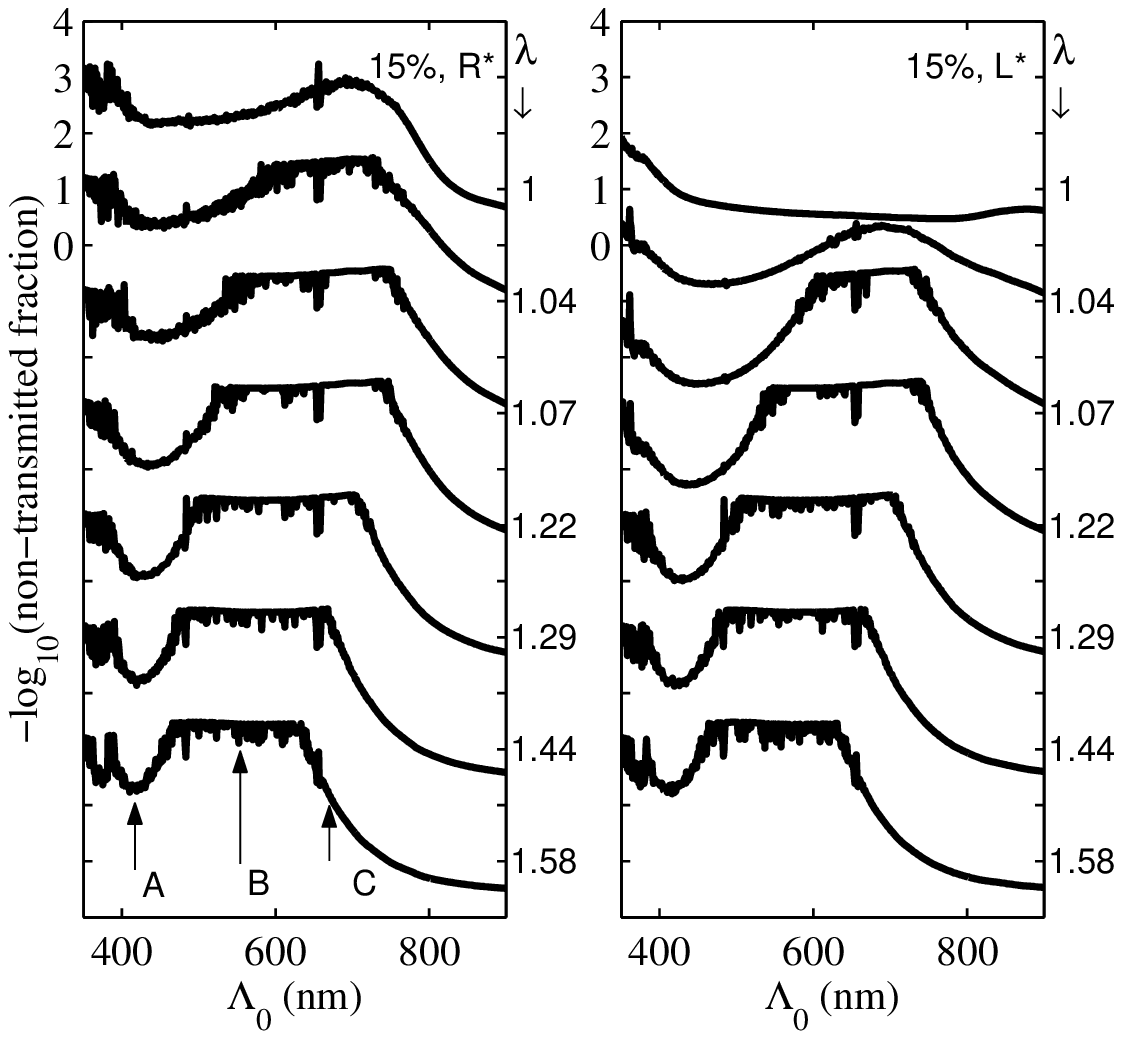}}\,\,\,\,\,
\resizebox{0.43\textwidth}{!}{\includegraphics{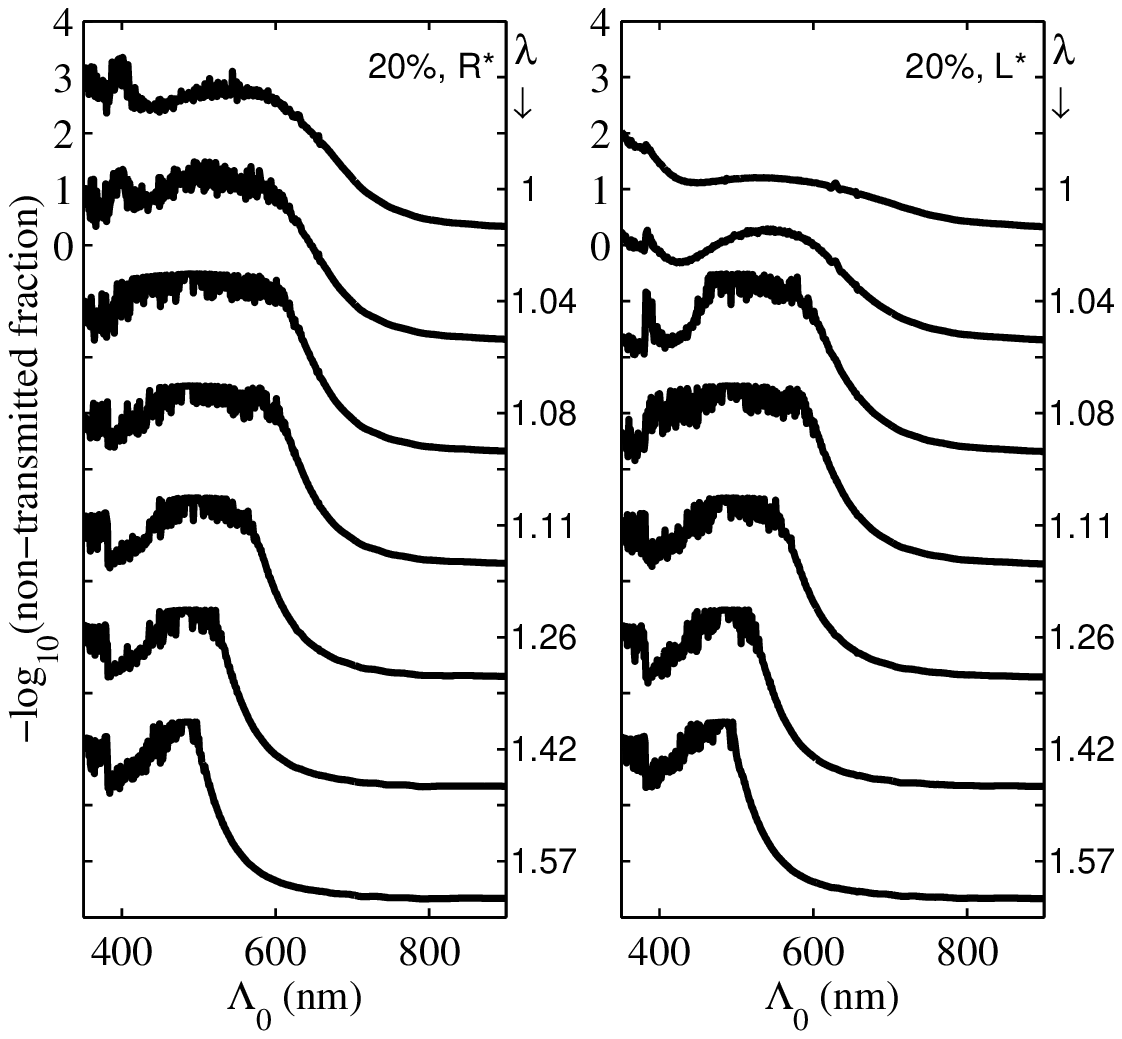}}\\
\resizebox{0.43\textwidth}{!}{\includegraphics{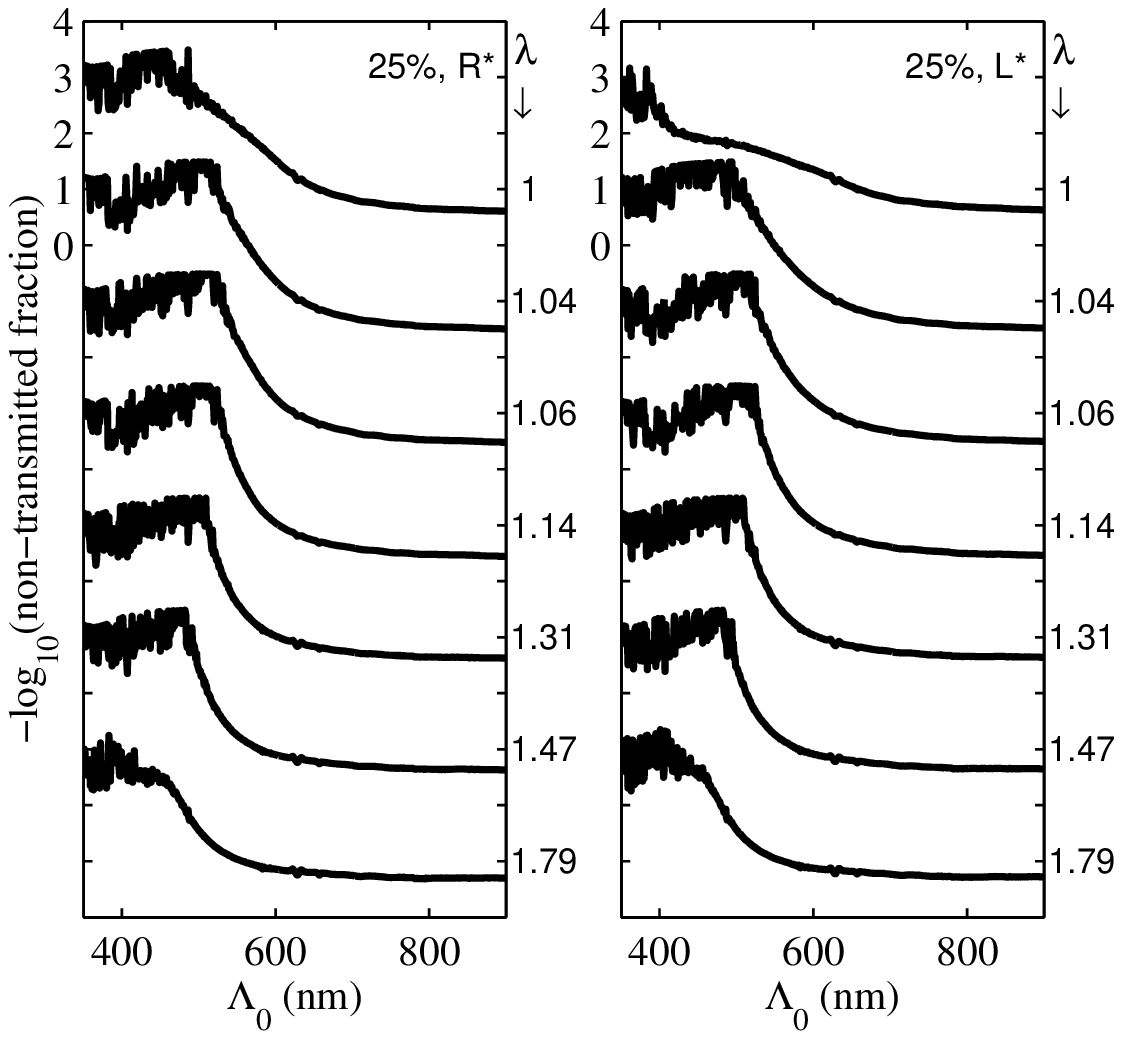}}\,\,\,\,\,
\resizebox{0.43\textwidth}{!}{\includegraphics{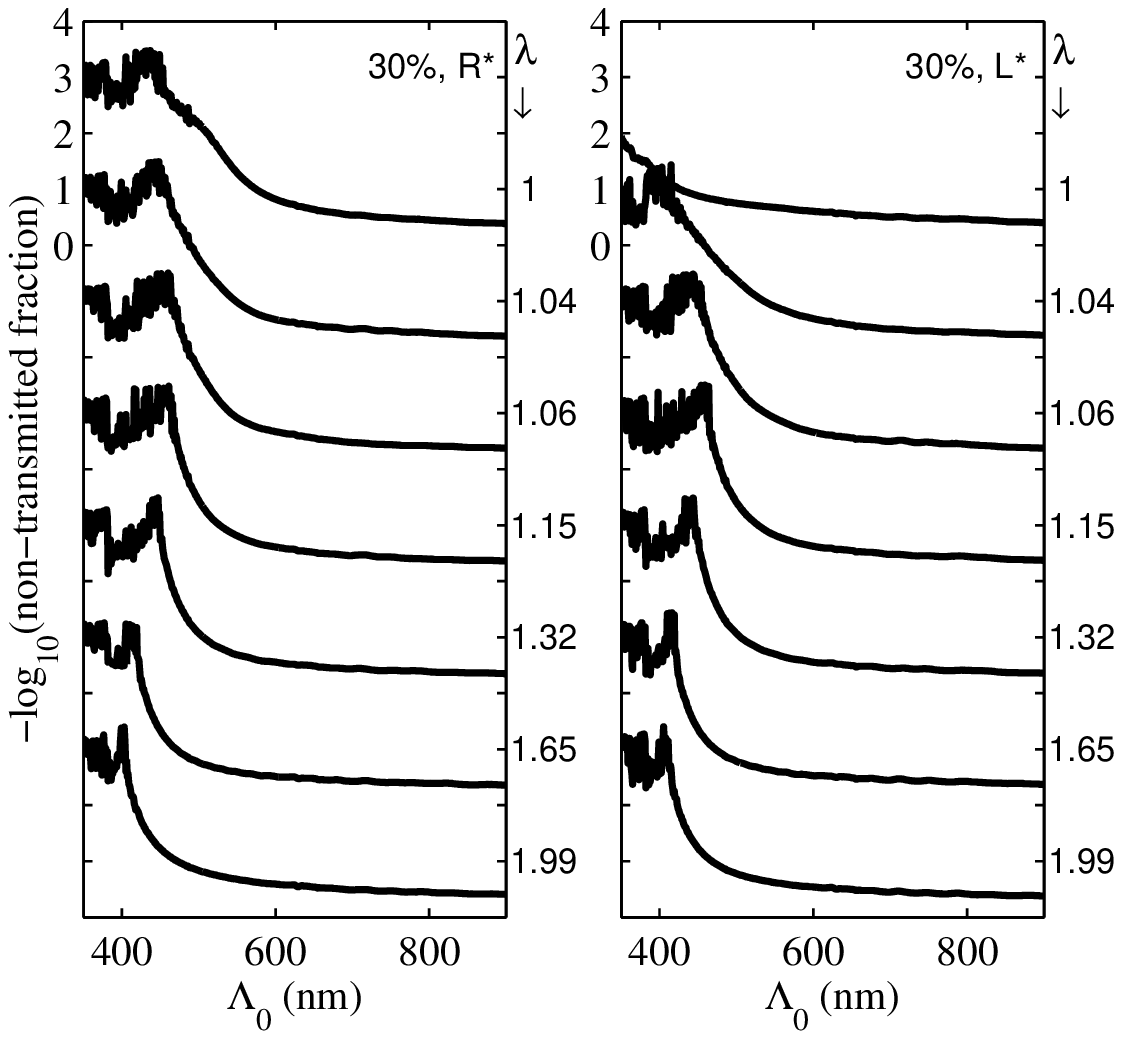}}
\caption{Transmittance spectra for opposite circular polarizations
through cholesteric elastomers with different fraction of chiral
dopant: 15\%, 20\%, 25\% and 30\%. The circular polarizations and
the concentration of chiral component are labelled in each panel.
Each material is uniaxially strained by the amount
$\lambda_{xx}=\lambda$, annotated beside the corresponding
spectrum. Successive spectra are shifted downwards by 2~absorbance
units for clarity. Flat regions of very low transmittance
correspond to the detection sensitivity limit of the spectrometer.
Three characteristic points are shown on the top-left panel and
their evolution is discussed in the text. A:~marks the minimum at
low wavelength; B:~marks the center of the reflection band;
C:~marks the point where the transmitted fraction is 0.01.
\label{spectra}}
\end{figure*}

Materials with four different concentrations of chiral groups have
been obtained, as shown in Fig.~\ref{chem}(a), with the molecular
proportions 15, 20, 25 and 30\%. The hydrosilation reaction, with
a platinum catalyst (from Wacker Chemie), was carried out under
centrifugation at 7000rpm initially for 25 minutes at 75$^\circ$C
to form a partially crosslinked gel. For the further 8 hours the
reaction proceeded under centrifugation at 30$^\circ$C, during
which time the solvent was allowed to evaporate, leading to an
anisotropic deswelling of the gel and completion of crosslinking.
All of the volume change in this centrifuge setup occurs by
reducing the thickness of the gel, while keeping the lateral
dimensions fixed. At the second-stage temperature of 30$^\circ$C
the dried gel is in the cholesteric phase and its director is
forced to remain in the plane of stretching -- this results in the
uniform cholesteric texture sketched in Fig.~\ref{diagram}. This
kind of synthesis produces a very homogeneous strip of elastomer
with the size of the order of 20cm$\times$1cm$\times$230$\mu$m.

Differential scanning calorimetry (DSC) was used to characterize
the phase sequence of resulting elastomers (Perkin-Elmer Pyris 7
DSC). Figure~\ref{chem}(b) shows the typical scans on heating,
indicating the glass transition temperature $T_g \approx
-7^\circ$C, as is common for polysiloxane side-chain polymers. The
clearing point, i.e. the isotropic-cholesteric transition
temperature, clearly depends on the composition, decreasing from
$T_{\rm c}\approx 65^\circ$C for {\em X}=15\%, to $T_{\rm
c}\approx 60^\circ$C for 20\%, $T_{\rm c}\approx 52^\circ$C for
25\% and to $T_{\rm c}\approx 46^\circ$C for 30\%. No additional
phase transformations are found between $T_{\rm c}$ and $T_g$. All
optical experiments are performed at room temperature,
sufficiently far from both transitions.

\subsection{Experimental technique}

Measurements  of the transmitted fraction of circularly polarized
light normally incident on the sample are made for wavelengths
across the visible spectrum. A combination of a linear polarizer
and a Fresnel rhomb is inserted in the optical path of a HP-8453
UV/Vis spectrophotometer, so that the incident light reaches the
sample having been circularly polarized. The clockwise R* or
anticlockwise L* handedness of circular polarization is determined
by the orientation of the linear polarizer. Light with a
wavelength below $\Lambda_0\sim320$nm is not transmitted due to
absorbtion in both the polarizer and the glass prism. For
$\Lambda_0 >$320nm the cholesteric materials under study reflect
backwards the fraction of light that is not transmitted.

\section{Photonic Stop-gaps}\label{Photonic}

The panels in Fig.~\ref{spectra} show the fraction of light
transmitted by the cholesteric elastomers as a function of strain,
for light of each circular polarization. We label right and left
handed circularly polarized light by R* and L* respectively. These
spectra are a measure of the photonic stop-gaps, and show the
dramatic effect of mechanical deformation.

It is possible to make some observations which hold generally for
the four materials under study. At no external strain
($\lambda_{xx}=1$) there is a very strong circular dichroism, the
samples transmitting most of the L* component of light but
reflecting R* for  a range of wavelengths. The reflection peak, as
expected,  shifts into the blue (shorter cholesteric pitch) on
increasing the proportion of chiral component in the material.  As
soon as a uniaxial strain is imposed, the transmitted intensity
spectra for L* polarized light show a continuous development of a
new reflection gap, at the same wavelength as the intrinsic R*
peak. This dramatic change in the transmittance was first reported
in \cite{Cicuta:02} for another monodomain CLCE network. Now we
can to study this effect in much greater detail because our
materials (especially the 15\% and 20\%) have a photonic gap
centered at large wavelength. The development of a gap in the
``opposite'' circular polarization, L*, even at a small strain,
corresponds to the sudden onset of a bright coloring of the
material, see Fig.~\ref{diagram}. Indeed once the peak in L*
reflectance has fully developed, the sample  reflects practically
the totality of incident light at that wavelength. With further
increasing uniaxial strain, the reflection gaps become more narrow
and shift to lower wavelength. This corresponds to the change in
color seen in stretched CLCE.

Above $\lambda_{xx} \approx 1.1$ and above the R* and L* spectra
are hardly distinguishable. This suggests that there is no
macroscopic chirality left in the stretched cholesteric networks,
which are apparently no longer twisted in a helical, albeit
coarsened, fashion. We shall discuss this point below, making
connection with optical rotation data and the state of stretched
CLCE above the critical strain $\lambda_c$.

\subsection{Effect of concentration of chiral dopant}

\begin{figure} 
\resizebox{0.35\textwidth}{!}{\includegraphics{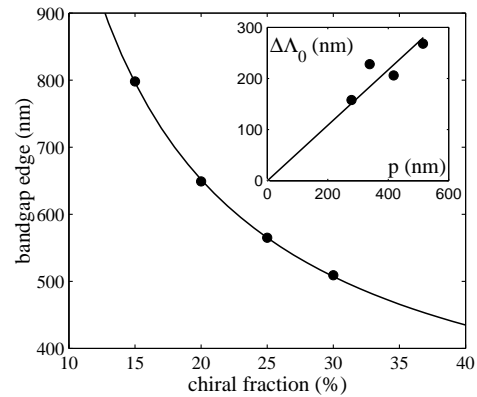}}
\caption{Position of the bandgap edge as a function of the
concentration {\em X} of chiral groups. The solid line is a fit to
the Eq.~(\ref{chiral}).The inset shows the approximately linear
dependence between the width of the bandgap, $\Delta p$, and the
helical pitch itself. Here the solid line is  a guide to the eye,
indicating the proportionality $\Delta \Lambda_0 \propto p$.
\label{posvsconc}}
\end{figure}

Figure~\ref{posvsconc} shows the position of the stop-band edge as
a function of the  concentration of chiral moieties in equilibrium
(at $\lambda=1$). As expected, the  helix pitch $p$ is tighter the
higher the concentration {\em X} of chiral component. We find an
excellent agreement with a relation
\begin{equation}
p\,=\, p_0\,+\,c/X   \label{chiral},
\end{equation}
indicating that for low {\em X}  each chiral molecular moiety
induces  chiral activity independently. For high {\em X}, the
pitch tends to a saturation value $p_0$, which is a reflection of
intrinsic chirality of the 2MOB${}^3$ molecular group. For our
materials, the fitting in Fig.~\ref{posvsconc} gives the values
$p_0=218$nm and $c=8674$nm (or $86.7$nm if {\em X} is more
naturally expressed as a fraction, instead of a percentage). From
this one could propose that if all side chains of our polymers
were made of the chiral 2MOB${}^3$-group, ${X}=100\%$, the
resulting cholesteric would have a pitch of $p \approx 295$nm.

Figure~\ref{posvsconc} also shows the width of the stop-band as a
function of the helical pitch wavelength. From the treatment of
light propagation in an ideal cholesteric helix \cite{deGennes93},
a proportionality is expected between the bandgap width $\Delta
\Lambda_0$ and the pitch:
 $$ \Delta \Lambda_0 = p \frac{m_{\rm e} - m_{\rm o}}{\bar{m}} .
 $$
For the refractive indices $m_{\rm e} \simeq 1.75$ and $m_{\rm o}
\simeq 1.6$ typical of thermotropic liquid crystals, the
proportionality reads $ \Delta \Lambda_0 \simeq 0.1 \, p$. For the
CLCE materials we observe the expected proportionality, but the
inset in Fig.~\ref{posvsconc} shows that the gap width is about
50\% of the pitch, at least five times  wider than in
``classical'' cholesteric liquid crystals. In addition, our
separate measurements (not shown) of the evolution of transmission
spectra with temperature (and, in consequence, with the nematic
order parameter $Q \propto m_{\rm e} - m_{\rm o}$) have indicated
that the gap position and width change very little as one
approaches the isotropic phase (although the amplitude of
selective reflection does, of course, diminish). This means very
likely that the pitch is not precisely single-valued throughout
the elastomer thickness and the band width is determined by the
quenched disorder inherent in the CLCE helix.

\subsection{Effect of strain}

\begin{figure} [t]
\resizebox{0.46\textwidth}{!}{\includegraphics{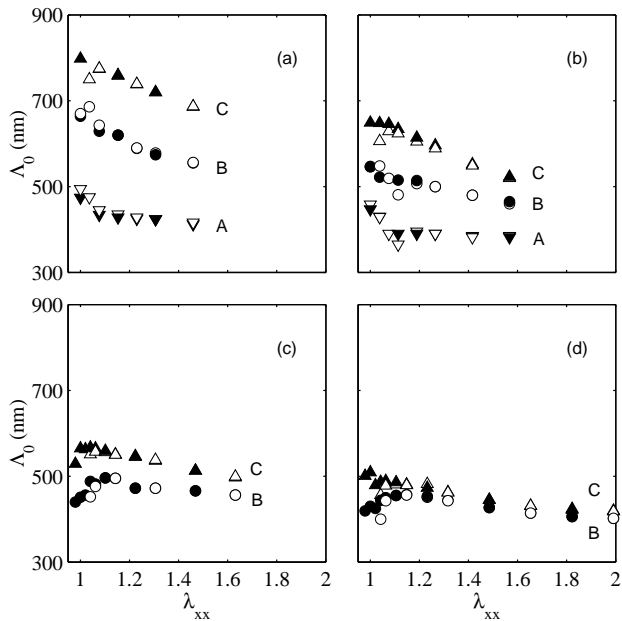}}
\caption{The effect of uniaxial strain on the photonic stop-gap.
A:($\blacktriangledown$), B:($\bullet$) and C:($\blacktriangle$),
 correspond to the characteristic positions on the reflection
peaks shown in Fig.~\ref{spectra}. Solid symbols correspond to R*
circular polarization and empty symbols to L*. The data are
obtained from materials with different chiral dopant concentration
{\em X}: (a)~15\%, (b)~20\%, (c)~25\% and (d)~30\%.
\label{gapmove}}
\end{figure}

Three characteristic  points can used to quantify the spectra
shown in Fig.~\ref{spectra}: the peak position (measured as the
center of the bandgap when the peak is at the detector
saturation), the long wavelength band edge (measured at the half
height of the reflection peak) and the center position of the
trough between the primary reflection peak and the next rise at
short (UV) wavelengths (only visible in 15 and 20\% materials).
The movement of each of these points is followed, as function of
imposed strain, in Fig.~\ref{gapmove}. The plots show data for
both the R* and the L* peaks but, of course, at very small strains
$\lambda < 1.1$ the L* reflection peak has not yet developed to
match the R* gap. For large strains the gaps are
indistinguishable, as the filled and open symbols in the plots
indicate, and shift together towards smaller wavelengths.

It is also clear that the gap becomes narrower as a function of
strain. There are two factors contributing to this. First of all,
as we have seen in the previous section, the gap width is
proportional to the pitch -- which is affinely decreasing with the
thickness of elastomer film (the middle data set in each of the
four plots in Fig.~\ref{gapmove}). Secondly, and perhaps more
importantly, the stretching of a cholesteric rubber film leads to
some reduction in effect due to the quenched disorder,
representing itself in the inhomogeneity of the local pitch as
well as the wavering of its axis about $z$. The increasing
external strain leads to greater alignment of the director in the
sample plane and, as a result, to a greater perfection of the
periodic texture. This is also the reason for increasing clarity
of colors reflected from the material.

\begin{figure} 
\resizebox{0.35\textwidth}{!}{\includegraphics{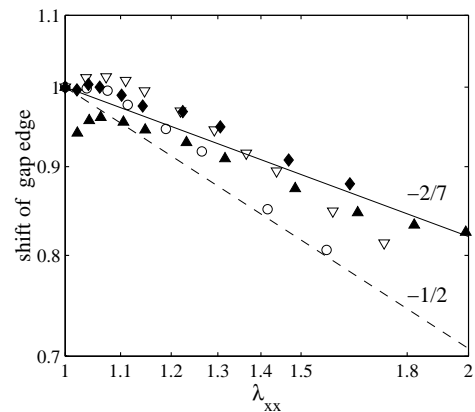}}
\caption{Plot of the  bandgap edge position relative to the value
for the undeformed material, as function of strain, on a
logarithmic scale. Symbols refer to materials with a different
concentration  of chiral dopant.~($\blacktriangle$):X=30\%,
($\blacklozenge$):X=25\%, ($\circ$):X=20\% and
($\triangledown$):X=15\%.  The lines correspond to two limiting
scaling laws as described in the text. \label{centergapscaling}}
\end{figure}

The relative movement of the R* reflection gap edges is shown in
Fig.~\ref{centergapscaling}. The pitch of the cholesteric helix,
which is probed by the lack of transmission of R* light, is
affinely deformed by the sample contraction along $z$-axis, and
the shift of the reflection gap edge is a direct measure of this
contraction, $\lambda_{zz}$. In an incompressible elastomer, an
imposed extensional strain $\lambda_{xx}$ induces a contraction in
$y$ and $z$ directions. Figure~\ref{centergapscaling} shows the
scaling of the gap position as a function of strain.  For an
isotropic rubber one would would always find a symmetric
relationship: $\lambda_{yy}= \lambda_{zz}= \lambda^{-1/2}$. It can
be clearly seen that the 30\% and 25\% samples follow a different
scaling, the $\lambda_{zz}$ strain following the power law
$\lambda_{xx}^{-2/7}$. This is an expected result, predicted by
semisoft elasticity of liquid crystal rubbers \cite{WT03}. It
occurs if the nematic director is allowed to rotate, as described
by Eq.~(\ref{phi}). Then, as explained in
Refs.~\cite{Warner:00,Bermel:02a}, the rubber strip is effectively
stiffer along the pitch axis and contracts much more in the plane
of the director (along $y$-axis) and much less in its thickness:
$\lambda_{yy}= \lambda^{-5/7}$ and $\lambda_{zz}= \lambda^{-2/7}$.

However the 20\% and 15\% samples appear to follow the well known
curve describing an isotropic rubber. We believe that this is due
to an increased effect of disorder in the alignment of the
director in the materials with a low chiral doping. The
non-uniform alignment of helical pitch would also reflect in the
large width of the bandgap as compared to an ideal helix. At
present these effects are not considered in the theoretical
description.

Additional bandgaps have been predicted
\cite{Bermel:02a,Bermel:02} to form under strain, at wavelengths
which are around half of the original helical pitch. Their
observation is difficult because they should appear close to our
experimental limit of low wavelength, even for the 15\% sample.

\section{Conclusions}

We have presented experimental results showing the very rich
optical behavior of monodomain cholesteric elastomers and studied
the effects induced by mechanical deformation.

Such materials can be synthesized with optical stop-bands anywhere
across the visible range. In analogy with a classical cholesteric
liquid crystal, the undeformed samples exhibit a very strong
circular dichroism, corresponding to a reflection band for light
of circular polarization R*, of the same helicity as the
cholesteric structure itself. As soon as a uniaxial strain is
imposed, a stop-band appears also for the opposite circular
polarization L* and the sample becomes an effectively
one-dimensional broadband Bragg-grating, reflecting all light. The
wawelength of the stop-gap is controlled by the external
deformation. These properties were theoretically predicted in
\cite{Warner:00,Bermel:02a} for an ideal material. However the
extraordinary large width of the reflection band in all materials
studied so far is a clear indication of imperfections in rubbery
cholesteric networks. We believe a significant amount of quenched
random disorder is inherent in these monodomain CLCE due to the
preparation procedure, showing itself in both local variations of
initial helical pitch $p$ and the wavering of its axis. Both
effects become suppressed on initial mechanical stretching in the
plane of the director. This is in contrast with almost defect-free
cholesteric networks that are prepared by densely crosslinking
very thin cholesteric films spontaneously aligned by rubbed
surfaces, such as those used in recent studies of low-threshold
lasing \cite{Schmidtke:03}.

To summarize, the present study confirms our understanding of the
effects of a mechanical field on a cholesteric liquid crystal
elastomer. These materials can be readily fabricated in large
strips and have already found applications. Their surprisingly
rich and well understood structural behavior will surely prove
useful in novel photonic devices. Further theoretical
investigation is required to describe the effects of quenched
random disorder on the modulation of phase of the optical axis.

\begin{acknowledgments}
We thank  Mark Warner for  a number of useful discussions. This
research has been supported by EPSRC.
\end{acknowledgments}

\bibliography{Refs}
\end{document}